\documentstyle[epsf]{laa}
\epsfverbosetrue
\newcommand{\eqb}{\begin{eqnarray}}
\newcommand{\eqe}{\end{eqnarray}}
\newcommand{\re}{{\cal E}}
\newcommand{\rp}{{\cal P}}
\newcommand{\wi}{{\cal I}}
\newcommand{\gesim}{\,\raisebox{-0.4ex}{$\stackrel{>}{\scriptstyle\sim}$}\,}

\def\mathrm#1{{\rm #1}}
\begin{document}
\thesaurus{02.01.1; 02.16.1; 02.19.1; 09.03.2; 09.19.2; 13.07.3}
\title{Limits on diffusive shock acceleration in dense and incompletely ionised media}
\author{L O'C Drury\inst{1} \and P Duffy\inst{2} \and J G Kirk\inst{2}}
\institute{Dublin Institute for Advanced Studies, 5 Merrion 
Square, Dublin 2, Ireland
\and
Max-Planck-Institut f\"ur Kernphysik, Postfach 10 39 80, D-69029 Heidelberg, Germany}
\offprints{L O'C Drury}
\date{Received August 21, 1995; Accepted October 12, 1995}

\maketitle

\begin{abstract}
The limits imposed on diffusive shock acceleration by upstream ion-neutral Alfv\'en 
wave damping, and by ionisation and Coulomb losses of low energy particles, are
calculated. Analytic solutions are given for the steady upstream wave excitation 
problem with ion-neutral damping and the resulting escaping upstream flux calculated.
The time dependent problem is discussed and numerical solutions presented.
Finally the significance of these results for possible observational tests of 
shock acceleration in supernova remnants is discussed.
\keywords{Acceleration of particles -- plasmas -- shock waves --
cosmic rays -- ISM: supernova remnants -- gamma rays: theory}
\end{abstract}

\section{Introduction}

Diffusive shock acceleration is generally believed to be an important
astrophysical mechanism for producing high energy charged particles
(Blandford \& Eichler~\cite{BE87}; Berezhko \& Krymsky~\cite{BK88};
Jones \& Ellison~\cite{JE91}). In particular, operating at the strong
shocks of Galactic supernova remnants, 
it is thought to be responsible
for producing the Galactic cosmic rays 
i.e., those of energy
less than about $10^{14}\,$eV. 
However this hypothesis has
not been convincingly confirmed by any direct observational test. Possibilities
which have been suggested include the observation of high energy
gamma-ray emission from supernova remnants, or detection of shock
precursor structures via Balmer line observations
(Aharonian et al.~\cite{aharoniandruryvoelk94}; Raymond~\cite{raymond91}).

A key element in the effective operation of the acceleration process
is the resonant excitation of scattering waves by the accelerated
particle pressure gradient ahead of the accelerating shock resulting
in much smaller values of the particle diffusion coefficient near the
shock than in the general medium. This was first emphasised by Bell
(\cite{bell78}) who gave an analysis of the process in the steady state
and also pointed out the importance of ion-neutral friction in damping
the waves and quenching the acceleration of high energy particles.

Another important aspect of diffusive shock acceleration is that it
can only operate in conjunction with an injection
mechanism, which must accelerate particles directly 
out of the `thermal' plasma up to a velocity of several times
the thermal speed. However in dense media low
energy injected particles are subject to Coulomb and ionisation
energy losses. The acceleration process has to be fast enough at
energies somewhat above `thermal' to compete with these collisional
processes, otherwise 
the shock will only be able to accelerate the few ambient
pre-existing high energy particles. 

These effects obviously have important implications for observational
diagnostics such as the gamma ray luminosity of supernova remnants
and, in the case of non-radiative shocks, the detailed line profiles
of the faint Balmer line emission. The gamma-ray emission will only be
easily detected if the shock is propagating into a relatively dense
medium with a number density in excess of $0.1\, \mathrm{cm}^{-3}$
(Aharonian et al.~\cite{aharoniandruryvoelk94}) and the Balmer diagnostics 
require that the medium
into which the shock is propagating be only about 90\% ionised
(Raymond~\cite{raymond91}). Rather generally one can anticipate that most
observational tests for cosmic ray acceleration in supernova remnants
will require some significant matter density to yield a detectable
signal. However too high a density may quench the acceleration
and destroy the very effect one is trying to detect.

In this paper we first develop the theory of wave damping and
excitation in cosmic ray shocks in more detail than has
been done previously and derive corrected estimates for the energy at
which this process quenches the acceleration. Then we consider
the conditions under which collisional processes can suppress the
natural shock injection process. Finally we discuss the
implications of these results for the observability of cosmic ray
acceleration effects in supernova remnants.

\section{Wave excitation and damping}

The model we study is identical to that considered by Bell
although we use a slightly different notation.  We ignore the reaction
of the accelerated particles on the flow, and consider only their
reaction on the scattering waves, which we take to be Alfv\'en waves,
moving at speed $V$ along the magnetic field
relative to a steady upstream flow of speed $U$
and density $\rho$ in the shock rest frame. Transmission of the waves through 
the shock and their damping downstream has been considered by 
Achterberg and Blandford~\cite{achterbergblandford86} and is not addressed here.
For simplicity, we will consider the case of parallel shocks, in which
the magnetic field lies along the normal to the shock surface.
Particles of momentum
$p$, charge $e$ and pitch $\mu$ ($\mu=\cos\theta$ where $\theta$ is
the angle between the particle's momentum vector and the mean magnetic
field direction) resonantly interact with waves of spatial wavenumber
along the field $k=1/(\mu r_{\rm g})$ where $r_{\rm g} = p/(eB)$ is the gyroradius
of the particle in the field of strength $B$ (there is an implicit
assumption here, that the waves are `slow' and the particles
`fast'). Although there is a contribution to the scattering of
particles of momentum $p$ by waves on all spatial scales smaller than
the particle gyroradius, it is clear that the bulk of the scattering
is from waves close to $r_{\rm g}$ in scale. In common with most discussions
of this topic we assume that particles of momentum
$p$, which from now on we take to mean particles in a (natural)
logarithmic interval of momentum, interact only with waves in a
logarithmic interval of $k$ space where $k$ and $p$ are implicitly
related by $kp=O(eB)$. This is often called
``sharpening the resonance.''

\section{Equations of the model}

The accelerated particles diffuse against the flow so that the spatial
and temporal evolution of the isotropic part of the particle phase
space density $f(t, x, p)$ is given by the advection-diffusion
equation 
\eqb
{\partial f\over\partial t} + (U-V){\partial f\over\partial x} =
{\partial\over\partial x}\left(\kappa {\partial f\over\partial
x}\right)
\eqe
with a diffusion coefficient $\kappa$
which we take, from quasi-linear theory, to have the form
\eqb
\kappa = {\kappa_{\rm B}\over\wi},\qquad \kappa_{\rm B} = {r_{\rm g} v\over 3}
\eqe
where $\wi$ is the dimensionless resonant wave intensity per
logarithmic interval of wavenumber and $\kappa_{\rm B}$ is the so-called
Bohm diffusion coefficient corresponding to a mean free path of order
the particle gyroradius. The total wave energy density is given by
\eqb
{{\left<\delta B^2\right>}\over{2\mu_0}}
={{\left<B\right>^2}\over{2\mu_0}}\int\wi(k)\,d\ln(k).
\eqe
Physically all this says is that the larger the fluctuations $\delta
B$ imposed on the mean field $B$, the stronger the scattering, until
when $\delta B\approx B$ the field is totally disordered
and particle trajectories lose coherence on scales as short as the
gyroradius.  Note that the advection is with the waves which we assume
to be moving backwards at velocity $V$ relative to the background
plasma, and thus to be moving at velocity $U-V$ relative to the shock
(which we locate at $x=0$).

In scattering off the waves the particles do work at the
rate $V\nabla P$ where $\nabla P$ is the accelerated particle
pressure gradient. The resonance sharpening simplification allows us
to apply this to a particular set of resonant particles and waves to
give the wave energy equation,
\eqb
{\partial I\over\partial t} + (U-V) {\partial I\over\partial x}
= V{\partial P\over\partial x} - \gamma I
\eqe
where $I$ is the (dimensional) wave energy density, $P$ the
resonant particle pressure and $\gamma$ a damping coefficient.
It is convenient to write this in
nondimensional form, and scale $I$ to the background field energy density
$B^2/2\mu_0=\rho_{\rm i} V^2/2$ and $P$ to the ram pressure of the background
flow of ionised plasma, $\rho_{\rm i} U^2$, where $\rho_{\rm i}$ is
the ion mass density.
We let $\rp$ denote the dimensionless pressure per logarithmic momentum 
interval,
\eqb
\rp = {{4\pi p^3}\over{\rho_{\rm i} U^2}} {{pv}\over{3}} f(p)
\eqe
so that the total accelerated particle pressure is
\eqb
\int{{pv}\over3}4\pi p^2f(p)\,dp
=\rho_{\rm i} U^2\int\rp(p)\,d\ln(p)
\eqe
Expressed in terms of the dimensionless quantities $\rp$ and $\wi$ we
then obtain the fundamental pair of coupled equations
\eqb
{{\partial \rp}\over{\partial t}} + (U-V) {{\partial\rp}\over{\partial x}} =
{\partial\over{\partial x}}\left( {\kappa_{\rm B}\over\wi} 
{{\partial\rp}\over{\partial x}}\right) 
\label{particleeqn}
\eqe
and
\eqb
{{\partial\wi}\over{\partial t}} +(U-V) {{\partial\wi}\over{\partial x}} =
{{2U^2}\over V}{{\partial\rp}\over{\partial x}} - \gamma \wi.
\label{waveeqn}
\eqe

We assume that the main process contributing to the wave
damping is ion neutral friction.  If the neutrals do
not move coherently with the ions, the resulting wave damping
rate $\gamma$ has the value
\eqb
\gamma = n_{\rm n}\left<\sigma v\right>
\eqe
where $n_{\rm n}$ is the neutral number density and $\left<\sigma v\right>$ is the
mean of the collision velocity times the charge exchange cross
section; Kulsrud \& Cesarsky~( \cite{KC71}) give the approximation
\eqb
\left<\sigma v\right> \approx 8.4\times 10^{-9}\left(T\over 10^4 \,\mathrm
K\right)^{0.4} \,\mathrm{ cm^{3}s^{-1}}
\eqe
for temperatures $T$ in the range $10^2\,\mathrm K$ to $10^5\,\mathrm
K$. The assumption that the neutrals do not participate in the
coherent oscillations of the ions in the Alfv\'en wave implies that
the period of the wave must be short relative to the momentum transfer
time scale from the ions to the neutrals. By assumption the wavelength
is of order the gyroradius of a resonant particle, thus this condition
implies 
\eqb
{r_{\rm g}\over V} < {1\over n_{\mathrm i}\left<\sigma v\right>}
\eqe
where $n_{\mathrm i}$ is the ion number density. Numerically, this
condition translates, for accelerated protons of energy $E$, to
\eqb
{E\over\mathrm{1\,GeV}} < 8 \left(T\over 10^4\,\mathrm
K\right)^{-0.4} \left(n_\mathrm{i}\over
1\,\mathrm{cm}^{-3}\right)^{-3/2} \left(B\over 1\,\mathrm{\mu G}\right)^2.
\label{dlimit}
\eqe
At higher energies the resonant waves are of sufficiently long period
for the neutrals to participate coherently in the wave motion. This
will lead to decreased wave damping and a reduction in the effective
Alfv\'en speed. The full dispersion relation for Alfv\'en waves in a
partially ionised medium is discussed by Kulsrud \& Pearce~(\cite{KP69}) 
and also by V\"olk et al.~(\cite{VMF81}). However for our purposes what 
is required is not
really the dispersion relation for freely propagating waves but the
damping of driven modes. An elementary calculation of the energy
dissipation in a frictionally coupled system of ions and neutrals when
the ions are forced at frequency $\omega$ gives
\eqb
\gamma = {\omega^2\over \omega^2 + \omega_{\rm i}^2} \omega_{\rm n}
\qquad \omega_{\rm i,n} = n_{\rm i,n}\left<\sigma v\right>
\eqe
Thus, as one would expect, at high frequencies the damping is constant
and equal to $\omega_{\rm n}$. However as $\omega \to \omega_{\rm i}$
the damping rate drops to $\omega_{\rm n}/2$ and at lower frequencies
decreases as $\omega^2$. Because the resonant wave frequency is
inversely proportional to particle momentum $p$ at sufficiently high
particle energies the damping rate of the resonant waves will decrease
as $p^{-2}$. 

\section{Parameters}

Looking at this system we see that it explicitly contains only two
dimensionless combinations of parameters. One is the ratio $V/U$ which
is the inverse Alfv\'en Mach number of the shock.  The other is
$\gamma\kappa_b/U^2$ which is essentially the ratio of the
acceleration time scale to the damping time scale. However the
boundary conditions on $\wi$ and $\rp$ introduce additional
dimensionless parameters into the problem.

The ratio $V/U$, for astrophysical shocks, is expected to lie in the
range $10^{-3}$ to 1.  In fact the assumption made above, that the
amplified waves are those propagating backwards relative to the flow
breaks down unless $V < U$.  For most of the calculations reported
here we have used $V/U=10^{-2}$. 

The dimensionless wave damping parameter due to ion-neutral friction
has the approximate numerical value at energies below the limit given 
by (\ref{dlimit}) 
\eqb
{\gamma\kappa_{\rm B}\over U^2} &=&
25.2 
\left(n_\mathrm{n}\over 1\,\mathrm{cm}^{-3}\right)
\left(T\over 10^4\,\mathrm K\right)^{0.4}
\left(E\over 1\,\mathrm{TeV}\right)\nonumber\\
&&\times\quad\left(B\over 1\,\mu\mathrm G\right)^{-1}
\left(U\over 10^3\,\mathrm{ km\,s^{-1}}\right)^{-2}
\eqe
where $n_\mathrm{n}$ is the neutral number density, $T$ the temperature and $E$
the accelerated particle energy (for relativistic protons).  Because
the acceleration time scale increases with energy the effects of wave
damping become more significant as the energy increases to the value
given by (\ref{dlimit}). At higher energies the damping rate decreases
faster than the acceleration rate. Thus the energy given by
(\ref{dlimit}) represents a critical threshold at which the wave
damping effects are strongest.  If the acceleration can reach this
energy, then the ion-neutral wave damping alone cannot restrict
further acceleration and the upper cut-off will be determined by other
factors such as shock geometry or age.

The dimensionless resonant particle pressure at the shock can be
estimated by assuming that the shock acceleration is efficient, so
that the total particle pressure is comparable to the ram pressure of
the flow, and that the spectrum is roughly that predicted by
test-particle theory for a strong shock, namely $f(p)\propto p^{-4}$ 
up to a maximum momentum $p_{\mathrm{max}}$ (which may be of order
$10^{14}\,\mathrm{eV}$ for supernova remnant shocks). This gives
\eqb
\rp \approx {1\over \ln(p_{\mathrm{max}}/mc)} \approx \hbox{0.1--0.05}
\eqe
as the (steady state) value at $x=0$, which is the right hand boundary condition
on $\rp$.  In general one could have a population of particles
advected into the shock from far upstream ($x \to -\infty$),
corresponding to a finite far upstream value of $\rp$, however the
physically interesting case is where all the particles are produced at
the shock so that the left hand boundary condition is $\rp \to 0$ as
$x\to -\infty$ (note that if $\wi$ also tends to zero this is compatible
with a nonvanishing escaping flux of particles upstream).

For the resonant wave intensity, we have to specify a left hand
boundary condition. One tempting possibility is to suppose that all
the waves are created by the resonant particles, so that $\wi \to 0$
as $x\to -\infty$.  However this is a rather singular limit as we will
see in the next section. In general one has to suppose that some
``seed'' field of waves exists in the far upstream plasma and is
advected into the shock where the backward propagating ones are then
amplified by the resonant particles.

\section{Steady analytic solutions}

If we set the time derivates to zero in Eqs.~(\ref{particleeqn}) and
(\ref{waveeqn}), these can be linearised by introducing a new 
independent variable $\tau$ similar to the 
\lq\lq optical depth\rq\rq\ variable used in 
radiative transfer:
\eqb
\tau&=&\int_{x_-}^x\,dx' {U-V\over\kappa(x')}
\label{optdeptheqn}
\\
&=&\int_{x_-}^x\,dx'\,\wi(x')\, {U-V\over\kappa_{\rm B}}
\nonumber
\enspace,
\eqe
where $x_-$ is an arbitrary point in the flow.  
We then find
\eqb
(U-V) {{d\rp}\over{d \tau}} &=&
{d\over{d \tau}}\left((U-V) 
{{d\rp}\over{d \tau}}\right) 
\label{newparticleeqn}
\eqe
and
\eqb
(U-V) {{d\wi}\over{d \tau}} &=&
{{2U^2}\over V}{{d\rp}\over{d \tau}} - 
{\gamma\kappa_{\rm B}\over U-V}
\label{newwaveeqn}
\enspace.
\eqe
Equation (\ref{newparticleeqn}) immediately yields a conserved quantity
-- the flux $\phi$ of particles:
\eqb
\phi&=&(U-V)\left(\rp-{d\rp\over d\tau}\right)
\eqe
and imposing the boundary condition $\rp=\rp_-$ at $x=x_-$ leads
to the solution
\eqb
\rp&=&
{\phi\over U-V}
+\left(\rp_- - {\phi\over U-V}
\right)
{\rm e}^\tau
\label{newparticlesol}
\enspace
\eqe
This may be substituted into Eq.~(\ref{newwaveeqn}), leading to 
the solution
\eqb
\wi&=&\wi_- -
{\kappa_{\rm B}\gamma\over(U-V)^2}\tau +
{2U^2
[\rp_- (U-V)-\phi]
\over V(U-V)^2}
\left({\rm e}^\tau - 1\right).
\label{newwavesol}
\eqe

Although it is in general both inconvenient and unnecessary to evaluate
$\rp$ and $\wi$ as functions of $x$, in the case of no damping
$\gamma=0$, the procedure is straightforward.
Defining $\alpha=2U^2[\rp_-(U-V)-\phi]/[(U-V)^2 V]$ and 
$\beta=(\alpha-\wi_-)(U-V)(x-x_-)/\kappa_{\rm B}$, one finds
\eqb
{\rm e}^\tau&=&{\alpha-\wi_-\over\alpha-\wi_-{\rm e}^\beta}
\eqe
so that
\eqb
\rp&=&{\phi\over U-V}+{\alpha V(U-V)(\alpha-\wi_-)\over 
2U^2
(\alpha-\wi_-{\rm e}^\beta)}
\eqe
and
\eqb
\wi&=&\wi_-{\rm e}^\beta\left({\alpha-\wi_-\over \alpha-\wi_-
{\rm e}^\beta}\right)
\enspace.
\eqe
These solutions diverge at a point $x_1$ given by
\eqb 
x_1&=&x_-+\kappa_{\rm B}\ln(\alpha/\wi_-)/[(\alpha-\wi_-)(U-V)]
\enspace,
\eqe
which must therefore be chosen to lie in the downstream region. In the 
special case $\alpha=0$, we recover the solution found by 
Lagage \& Cesarsky~(\cite{lagagecesarsky83}) by allowing 
$x_-\rightarrow-\infty$, whilst keeping $x_1$ finite:
\eqb
\wi&=&\wi_-\left\{1 - \exp\left[(U-V)\wi_-
(x-x_1)\over\kappa_{\rm B}\right]\right\}^{-1}
\enspace,
\label{lagagesoln}
\eqe     
which, in the limit $\wi_-\rightarrow0$, reduces     
to the somewhat simpler solution found by
Bell~(\cite{bell78}):
\eqb
\wi &=& {\kappa_{\rm B}\over (U-V)(x_1-x)}
\enspace.
\label{bellsoln}
\eqe
This is also a good approximation to the 
more general solution Eq.~(\ref{lagagesoln}) for $\wi\gg\wi_-$.
We remark at this point that the transformation Eq.~(\ref{optdeptheqn}) 
has eliminated the uninteresting solution $\rp=$constant, $\wi=0$. 
Nevertheless,
$\wi_-\rightarrow 0$ is a singular limit of the
full set (\ref{particleeqn}) and (\ref{waveeqn}).  

When damping is included, we note from Eq.~(\ref{newwavesol})
that there exists a point where the wave growth is just balanced 
by damping and $d\wi/d\tau=0$. 
Following Bell~(\cite{bell78}), we
assume that at this point the particles escape freely into the 
upstream plasma. 
As usual in the diffusion approximation, we 
can implement this boundary condition 
by demanding that the density $\rp$ vanish there.
Let us choose $x=x_-$ and $\tau=0$
at the free escape boundary.
Since the derivative of 
$\wi$ there is by definition zero and we
also demand $\rp_-=0$, the escaping particle flux is determined
independently of $\wi_-$ to be:
\eqb
\phi&=&-{\kappa_{\rm B}\gamma V\over 2 U^2}
\label{escflux}
\enspace.
\eqe
As pointed out by Bell, the escaping flux will steepen the spectrum if
it is comparable to the flux advected away downstream. Thus, wave
damping leads to a cut-off in the spectrum where
\eqb
{\gamma\kappa_{\rm B}\over U^2} {V\over 2} \approx {U-V\over4}\rp_0
\label{cutoff}
\eqe
where $\rp_0$ is the value of $\rp$ at the shock. Note however that
Bell~(\cite{bell78}), and following him Draine and
McKee~(\cite{DM}), overestimate the escaping flux by a factor
$c/(U-V)$ where $c$ is the speed of light {\it in vacuo}.
This error resulted from the use of the solution
given in Eq.~(\ref{bellsoln}), in which the particle flux is identically 
zero everywhere, as a basis for the estimate of the escaping flux.

\section{Stability and time dependence}

If we examine the system of equations and think about the physics
involved, it is clear that the wave excitation is strongest in those
regions which already have an enhanced level of wave activity, and
thus smaller values of the diffusion coefficient and steeper pressure
gradients.  It follows that the stability of the system is rather
questionable; there is a clear possibility of some form of
modulational instability. It is also possible that the steady
solutions are not in fact the physically realistic ones. Before
examining these questions numerically, it is interesting to look at
the case of small high-frequency perturbations, which can be examined
analytically.

We begin by linearising about a smooth background solution, $\bar\rp$
and $\bar\wi$, on which is superimposed a small fluctuating component,
$\tilde\rp, \tilde\wi$. The linearised equations are
\eqb
{D \tilde\rp\over Dt} = {\partial\over\partial x}
\left({\kappa_{\rm B}\over\bar\wi}{\partial\tilde\rp\over\partial x}
- {\kappa_{\rm B}\over\bar\wi^2}\tilde\wi{\partial\bar\rp\over\partial x}\right)
\eqe
and
\eqb
{D\tilde\wi\over Dt} = {2U^2\over V} {\partial\tilde\rp\over \partial
x} -\gamma \tilde\wi
\eqe
where 
\eqb
{D\over D t} = {\partial\over\partial t} + (U-V){\partial\over\partial
x}
\eqe
denotes the advective derivative.
We now assume that the fluctuations are of high spatial and temporal
frequency and make a formal expansion in inverse powers of the
wavenumber $k$,
\eqb
\tilde\rp &=&
e^{i\theta(x, t)}\sum_{n=0}^\infty k^{-n}\tilde\rp_n \\
\tilde\wi &=&
e^{i\theta(x, t)}\sum_{n=0}^\infty k^{-n}\tilde\wi_n
\eqe
where $\theta(x, t)$ is the rapidly varying phase,
$\partial\theta/\partial x = - k$ and $\partial\theta/\partial t =
\omega = O(k)$.
Substituting and collecting similar powers of $k$ we obtain, to order
$k^2$ and $k$,
\eqb
\tilde\rp_0 &=&  0 \\
\tilde\rp_1 &=& i {\tilde\wi_o\over\bar\wi} {\partial\bar\rp\over x} \\
{D\tilde\wi_0\over D t} &=& -i {2U^2\over V}\tilde\rp_1 -\gamma\tilde\wi_0
\eqe
together with the relation $\omega = (U-V) k$. Combining these results
we get
\eqb
{D \tilde\wi_0\over D t} =
{2U^2\over V} {\tilde\wi_0\over\bar\wi} {\partial\bar\rp\over\partial
x}-\gamma\tilde\wi_0 
\eqe
which we can write in the form
\eqb
{1\over\tilde\wi_0}{D\tilde\wi_0\over Dt} =
{1\over\bar\wi}{D\bar\wi\over Dt}.
\eqe

It follows that, to lowest order, small-scale perturbations in the
resonant wave intensity are advected in at velocity $U-V$ and grow (or
decay) at exactly the same rate as the general background wave
intensity. While this is not an instability, it does mean that the
system ``remembers'' all the fluctuations in the initial seed wave
field and is neutrally stable; in this lowest order linear high
frequency analysis the relative amplitude of perturbations neither
grows nor decays. Clearly a numerical study of non-linear finite size
perturbations is desirable.

There is no particular problem in developing a time-dependent
numerical scheme for solving this system of equations. However some
thought has first to be given to a time-dependent right hand boundary
condition for $\rp$.  The problem arises because we are trying to look
at one small part of the general process of shock acceleration in
isolation. Essentially we have to model the rest of the shock
acceleration process in a simple, but physically plausible way, in the
boundary conditions we use.  We assume that there is a steady flux,
of particles into the momentum interval we are considering
from acceleration at lower momenta.  We are not interested in
modelling the downstream scattering, so we simply assume that there is
efficient scattering downstream and that the accelerated particles
fill a downstream phase space ``reservoir'' of size $4\pi p^3 L$ with
$L\approx \kappa_{\rm B}/W_2$ where $W_2$ is the downstream advection
velocity (we avoid discussion of whether this is the plasma velocity
minus the Alfv\'en velocity). From this they can diffuse into the
upstream region and they can be removed either by advection
downstream, or by being accelerated to even higher energies.

We need an equation for the resonant particle pressure, $\rp$, at the
shock and in the downstream ``reservoir''.  The required equation is
simply energy conservation as applied to the reservoir.  If we denote
by $W_1 = U-V$ the advection velocity into the shock and by $\re$ the
energy density in resonant particles, then the energy flux in resonant
particles into the reservoir from upstream is $W_1(\re+\rp) -
\kappa(\partial
\re/\partial x)$, the flux out downstream is $W_2(\re+\rp)$, the flux
in from lower momenta is $\re_0(W_1-W_2)/3$ where $\re_0$ is the
energy density per logarithmic interval at lower energies and the flux
out to higher energies is $\re(W_1-W_2)/3$. Thus
\eqb
L {d\re\over dt} & = &
W_1(\re+\rp) - {\kappa_{\rm B}\over\wi}{\partial\re\over\partial x}\nonumber\\
&&- W_2(\re+\rp) -  {W_1-W_2\over 3}(\re - \re_0).
\eqe
Note that, because we are only considering particles in a specific
momentum interval, this is not a two-fluid approximation. To simplify
matters we assume that $W_2 = W_1/4$ and that we are dealing with
relativistic particles where $\re = 3\rp$. The right hand boundary
condition for $\rp$ then becomes simply
\eqb
L {d\rp\over dt} = W_2 (\rp_0 + 3\rp) -
{\kappa_{\rm B}\over\wi}{\partial\rp\over\partial x}
\label{rhsbc}
\eqe
where $\rp_0$ is the value to which $\rp$ tends if the solution
approaches the standard steady result (note that in a steady solution
\eqb
{\kappa_{\rm B}\over\wi}{\partial\rp\over\partial x} = 4 W_2 \rp).
\eqe

\section{Numerical solutions}

It is not hard to solve the fundamental system of equations
(\ref{particleeqn}, \ref{waveeqn}) numerically. We have used a uniform
grid and fixed time step with $\Delta x/\Delta t = (U-V)$ so that the
advection is exact. The diffusion is treated fully implicitly giving a
simple tridiagonal system and the wave intensity is defined on a
staggered mesh between each pair of $\rp$ values. The wave damping is
also treated implicitly so that the scheme is absolutely stable for
all diffusion coefficients and damping rates. For more elaborate
calculations there would be definite advantages in going to variable
grid spacings and time steps, and perhaps a second order scheme (our
scheme is only of first order). However for investigating stability
there are distinct advantages in using a simple first order scheme
with no advection errors. 

\begin{figure}
 \epsfxsize=\hsize
 \epsfbox{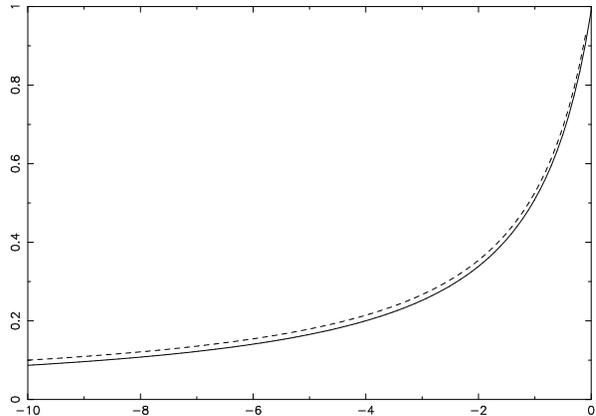}
 \caption{The numerical solution after time $100\kappa_{\rm B}/U^2$ for $V/U=0.01$,
$\gamma=0$ and $\rp_0=0.005$. The dashed line is $\wi$ and the solid line is
$\rp/\rp_0$. Distance upstream is in units of $\kappa_{\rm B}/U$.}
 \label{fig2}
\end{figure}

\begin{figure}
 \epsfxsize=\hsize
 \epsfbox{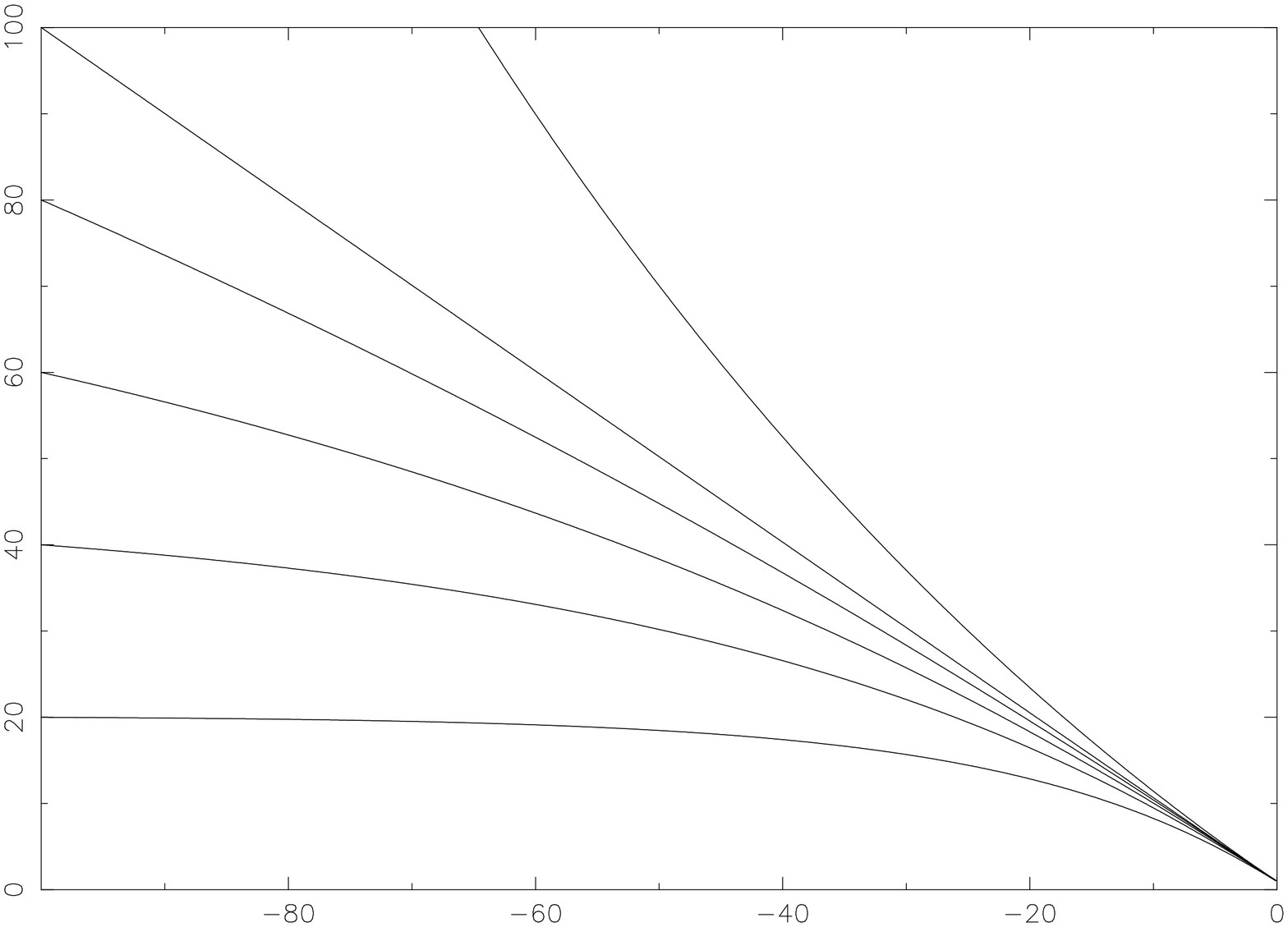}
 \caption{The reciprocal of the resonant wave intensity, $1/\wi$, 
          after time $200\kappa_{\rm B}/U^2$
          in the region from $-100\kappa_{\rm B}/U$ to 0. The parameters are as in  
          Fig.~\protect{\ref{fig2}}
          except that the the left-hand value of $\wi$ is successively set to 0.005,
          0.01, 0.0125, 0.01667, 0.025 and 0.05.}
 \label{fig3}
\end{figure}

We first demonstrate that this program can reproduce the analytic results
for steady solutions. In Figure \ref{fig2} we show the solution attained at time 100
(in units of $\kappa_b/U^2$) starting from $\rp=0$ and $\wi=0.1$. The
parameters have been chosen so that in Bell's solution (\ref{bellsoln})
the right-hand value of $\wi$ is unity. The left-hand boundary
conditions are zero net flux for $\rp$ and fixed $\wi=0.1$ for the
``seed'' waves.  In Fig \ref{fig3} we show $1/\wi$ over an extended range and
for various upstream seed values. Note that with an upstream value of 0.01 the
plot of $1/\wi$ is an almost perfect straight line, as required by Bell's
solution, and that the solutions for other seed values tend asymptotically to
this solution in the manner described by Lagage and Cesarsky's solution.

\begin{figure}
 \epsfxsize=\hsize
 \epsfbox{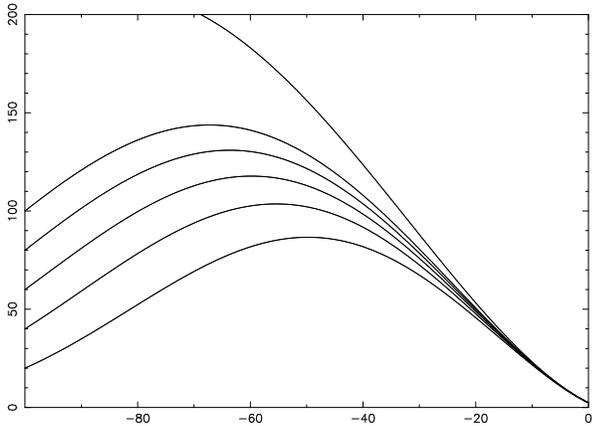}
 \caption{The reciprocal of the resonant wave intensity, $1/\wi$, with modest damping
          after time $200\kappa_{\rm B}/U^2$
          in the region from $-100\kappa_{\rm B}/U$ to 0. The parameters are as in 
          Fig.~\protect{\ref{fig3}} except that the damping rate $\gamma=0.1$.}
 \label{fig4}
\end{figure}

\begin{figure}
 \epsfxsize=\hsize
 \epsfbox{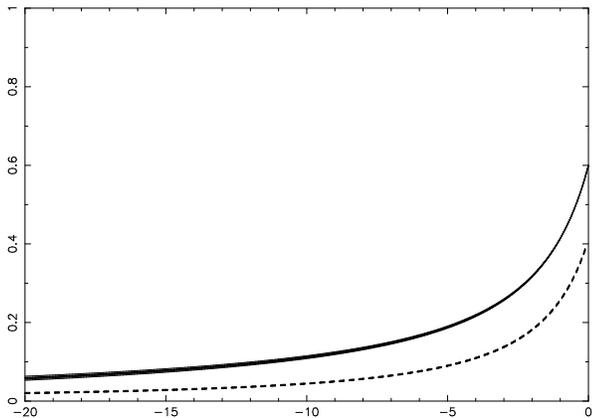}
 \caption{The resonant wave intensity $\wi$ (dashed lines) and $\rp/\rp_0$ 
(solid lines)
near the shock for the six solutions of Fig.~\protect{\ref{fig4}}. }
 \label{fig5}
\end{figure}

In figure Fig.~\ref{fig4} we show the effect of including modest damping on the
solutions of Fig.~\ref{fig3}. As is clear physically, and obvious from the analysis
of section 4, far upstream the damping dominates. However the various solutions do
converge to a single asymptotic solution near the shock. This is rather easier to
see in Fig.~\ref{fig5} where the final sections of the solutions of Fig.~\ref{fig4}
are plotted in the style of Fig.~\ref{fig2}. The left-hand boundary condition for
these solutions was taken to be the constant escaping flux given by equation 
\ref{escflux}. It is then easy to see that the right-hand boundary condition
(\ref{rhsbc}), in the steady state, implies
\eqb
\rp = \rp_0 -{\gamma\kappa_{\rm B}\over U^2}{V\over 2 W_2}
\eqe
which for the parameters used here gives $\rp = 0.003$ or $\rp/\rp_0 = 0.6$, as shown
by the numerical solutions. We note that this argument gives an approximate form for
the spectrum at energies below the cut-off given by equation (\ref{cutoff}).

\begin{figure}
 \epsfxsize=\hsize
 \epsfbox{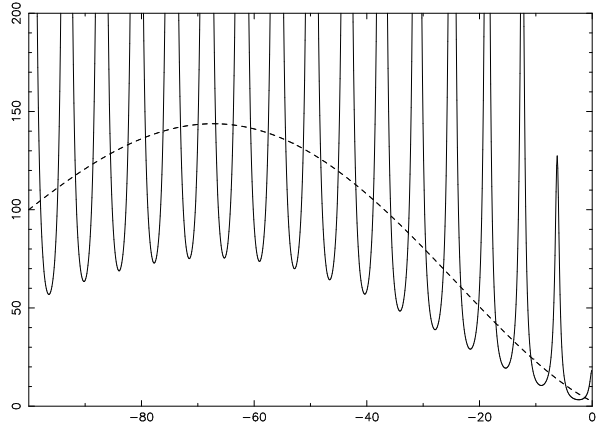}
 \caption{The reciprocal of the wave intensity with $V/U=0.01$,
 $\gamma=0.1$, $\rp_0=0.005$ after time $200\kappa_{\rm B}/U^2$ with
 a sinusoidally modulated and a steady left-hand boundary condition for
 $\wi$.  The solid curve represents the solution with $\wi=0.01(1+0.9\sin t)$
 as boundary condition, the dashed curve that with $\wi=0.01$.}
 \label{fig6}
\end{figure}

\begin{figure}
 \epsfxsize=\hsize
 \epsfbox{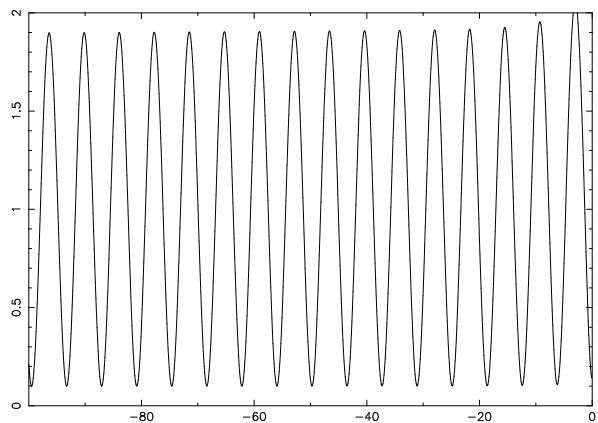}
 \caption{The ratio of the two solutions shown in Fig.~\protect{\ref{fig6}}.}
 \label{fig7}
\end{figure}

However it is most unlikely that the upstream ``seed'' field will be
absolutely steady.  In Fig.~\ref{fig6}
we show the solution from Fig.~\ref{fig4} corresponding to a steady left-hand
boundary condition of $\wi=0.01$ and also that of the strongly modulated
boundary condition, $\wi=0.01(1+0.9\sin t)$. More revealingly, in
Fig.~\ref{fig7}, we plot the ratio of these two solutions. It is
remarkable that, even at this very high level of modulation, the
solution behaves exactly as predicted by the analytic theory. Only
very close to the shock is there some evidence of a slight increase in the 
perturbation amplitude. It is also noteworthy that the steady solution
corresponding to the ``average'' value of $\wi$ is a good approximation
to the ``average'' time-dependent solution. Calculations with different periods and 
amplitudes yield similar results; although locally very different, the
large-scale structure appears to be well represented by the steady solution
corresponding to the ``average'' value of $\wi$.

In fact one has to be rather careful about exactly what sort of
average is meant in the above statements. If one is spatially
averaging along the particle gradient, the appropriate mean diffusion
coefficient is the harmonic mean, corresponding to the simple
arithmetic mean of the wave intensity. If one is averaging in the
perpendicular plane, the appropriate mean is the arithmetic mean of
the diffusion coefficient. More generally one has to consider the
geometry of the fluctuations as in percolation theory.  It may be
helpful to think of the electrical analogy of resistance networks
connected in series and in parallel.  To some extent, the fact that
the arithmetic mean of the wave intensity generates a steady solution
which is a very good approximation to the ``average'' non-steady
solution, is an artifact of the strictly one-dimensional model we are
using.

Although many extension or generalisations of diffusive shock acceleration
have been considered, as far as we are aware there has been no work
on shock acceleration with time-dependent scattering.  If one considers the
standard microscopic arguments leading to the steady-state spectrum for
test-particle acceleration, it is clear that the time-averaged spectrum at the
shock will be the same power-law spectrum, independent of the details of the
time-dependent scattering, as long as three conditions are met.
All particles heading upstream must return to the shock, which will be the
case for negligible wave damping or if there exists sufficient upstream
wave power. The angular distribution of the accelerated particles at the
shock front must be close to isotropic, which is
generally required for the diffusive
description to be applicable. And, crucially, the mean flux advected away downstream
must equal the downstream advection velocity times the mean density at the shock.
This last assumption appears open to question although it is in fact used in
our approximate right-hand boundary condition. This may be an interesting
way to produce non-standard spectral indices.

\section{Summary of wave damping effects}

In the preceding sections we have generalised the analytic theory of steady
resonant wave excitation to include wave damping and obtained an
analytic formula for the resulting escaping particle flux. In reality
the solutions will be time-dependent, but analytic arguments and
numerical experiments show that ``on average'' the steady solutions
can still give a good representation of the system (although this
would probably change if more complicated non-linear processes were
considered).  

The limit we derive for the maximum energy to which particles can be
accelerated before the escaping upstream flux kills the acceleration
differs substantially from that quoted in Draine and McKee~(\cite{DM}). 
Their formula (4.13) gives
\eqb
E_{\rm uG} < 3\times10^{-3}
{v_{\rm s7}^4 x_\mathrm{i}^{1/2} \rp_0\over (1-x_\mathrm{i})
n_0^{1/2}T_4^{0.4}}
\eqe
where in their notation $E_{\rm uG}$ is the upper cut-off energy for a
proton in units of GeV, $v_{\rm s7}$ is the shock speed in units of
$10^7\,\rm cm\,s^{-1}$, $n_0$ is the total number density and $x_i$ the
ionisation fraction. Note that their dimensionless particle pressure
parameter $\phi$ is related to ours by
\eqb
\phi = x_{\rm i} \int \rp_0 d\ln p
\eqe

Our result (\ref{cutoff}) says that if the 
upper cut-off energy $E$ is determined by
the ion-neutral wave damping in the upstream region, then numerically
\eqb
{E\over 1\mathrm{TeV}} &<& 
\left(U\over 10^3\,\mathrm {km\,s^{-1}}\right)^3
\left(T\over 10^4\,\mathrm K\right)^{-0.4}\nonumber\\
&\times&
\left(n_\mathrm{n}\over 1\,\mathrm {cm^{-3}}\right)^{-1}
\left(n_i\over 1\,\mathrm {cm^{-3}}\right)^{0.5}
\left(\rp_0\over 0.1\right)
\label{numcutoff}
\eqe
where $n_\mathrm{n}$ is the neutral number density and $n_i$ the ion density
in the medium into which the shock is propagating. 
Our limit is a factor $c/v_{\rm s} = 3\times 10^3/v_{\rm s7}$ times
higher. In fact, as discussed above, the damping will become unimportant above
energies given by (\ref{dlimit}). The condition that the upper cut-off
energy exceed that at which the neutrals start to coherently move with
the ions is
\eqb
{\rp_0\over 0.1}
\left(U\over 10^3\,\mathrm{km\, s^{-1}}\right)^3 
&>&
8\times 10^{-3} \left(n_\mathrm n\over 1\,\mathrm{cm}^{-3}\right)
\nonumber\\
&\times& \left(n_\mathrm i\over 1\,\mathrm{cm}^{-3}\right)^{-2}
\left(B\over 1\,\mathrm{\mu G}\right)^2.
\label{nolimit}
\eqe
If this condition is satisfied upstream ion-neutral wave damping places no
restriction on shock acceleration.

\section{Collisional losses near injection}

\newcommand{\Msolar}{M_{\odot}}
\newcommand{\coulrate}{t^{-1}_{\rm C}}
\newcommand{\ionrate}{t^{-1}_{\rm ion}}
\newcommand{\accrate}{t^{-1}_{\rm acc}}
\newcommand{\rcomp}{r}
\newcommand{\kBohm}{\kappa_{\rm B}}
\newcommand{\Tfour}{T_{4}}

At low energy, the most important loss processes for cosmic rays are
ionisation and Coulomb losses. If the theory of diffusive acceleration
at shocks is to be viable, it must overcome these for all energies
above that of injection. The constraints that this imposes on the
density, temperature and magnetic field of the ambient medium as well
as on the energy of injection can be estimated simply by comparing the
loss-rates with the acceleration rate.

Standard treatments of diffusive acceleration give the acceleration rate in
terms of the rate of change of momentum,
\eqb
\dot p&=&p{(u_1-u_2)\over3(\kappa_1/u_1+\kappa_2/u_2)}
\enspace,
\eqe
where the subscript refers to the upstream (1) or downstream (2)
medium, $\kappa$ is the diffusion coefficient and $u$ the 
fluid speed. However loss rates are normally discussed in terms of
the particle kinetic energy; defining the energy acceleration
rate $\accrate$ as the 
rate of change of kinetic energy divided by the kinetic energy we get
\eqb
\accrate&=&{u_1^2\over\kappa_1}
{\gamma\beta^2\over (\gamma-1)}
{(\rcomp-1)\over \rcomp
[1+(\rcomp\kappa_2/\kappa_1)]}
\enspace,
\eqe
where $\rcomp=u_1/u_2$ is the compression ratio of the shock, $\beta c$
is the speed of the particle and the Lorentz factor is $\gamma=(1-\beta^2)^{-1/2}$. 
The diffusion coefficient can be conveniently parameterised in terms
of the Bohm value in a magnetic field $B$:
\eqb
\kappa_1&=&k_1\kBohm\\
&=&k_1{\beta^2\gamma m_{\rm p}c^2\over3 eB}
\enspace,
\eqe
leading to 
\eqb
\accrate&=&1.1\times10^{-7}\, k_1^{-1}
\left(u_1\over1000\,{\rm km\,s^{-1}}\right)^2
\left(B\over 1\,\mathrm{\mu G}\right)
\nonumber\\
&\times&
\left[(\rcomp-1)\over\rcomp[1+(\rcomp k_2/k_1)]\right]
{1\over\gamma - 1}{\rm s^{-1}}
\enspace.
\eqe
At non-relativistic energies the acceleration rate is proportional to $\beta^{-2}$.

Expressions for the loss processes have been given by Mannheim \& 
Schlickeiser~\cite{mannheimschlickeiser94}. For Coulomb collisions
with free electrons of temperature $T=\Tfour\times10^4\,$K and density
$x_{\rm i} n_{\rm H}\,{\rm cm^{-3}}$, where $x_{\rm i}$ is the ionisation fraction and
$n_{\rm H}$ the total (ionised, neutral and molecular) density of hydrogen atoms, 
one finds
\eqb
\coulrate&=&3.3\times10^{-16} x_{\rm i} n_{\rm H}{\beta^2\over(\gamma-1)
(\beta_{\rm th}^3+\beta^3)}
\ {\rm s^{-1}}
\enspace,
\eqe
where $\beta_{\rm th}c$ is the thermal speed of the electrons: 
$\beta_{\rm th}=2\times10^{-3} \Tfour^{1/2}$. 
A similar expression is valid for losses due to
the ionisation of neutral material. Using a composition appropriate 
to the interstellar
medium, Mannheim \& Schlickeiser~(\cite{mannheimschlickeiser94}) find
\eqb
\ionrate&\approx&3.9\times10^{-16}(1-x_{\rm i})n_{\rm H}
\nonumber\\
&\times&
\left[
{\beta^2\over(\gamma-1)(\beta_0^3+2\beta^3)}
\right]
\ {\rm s^{-1}}
\enspace,
\eqe
where $\beta_0c$ is the speed of an electron in the ground state of the hydrogen
atom: $\beta_0\approx0.01$.

\begin{figure}
 \epsfxsize=\hsize
 \epsfbox{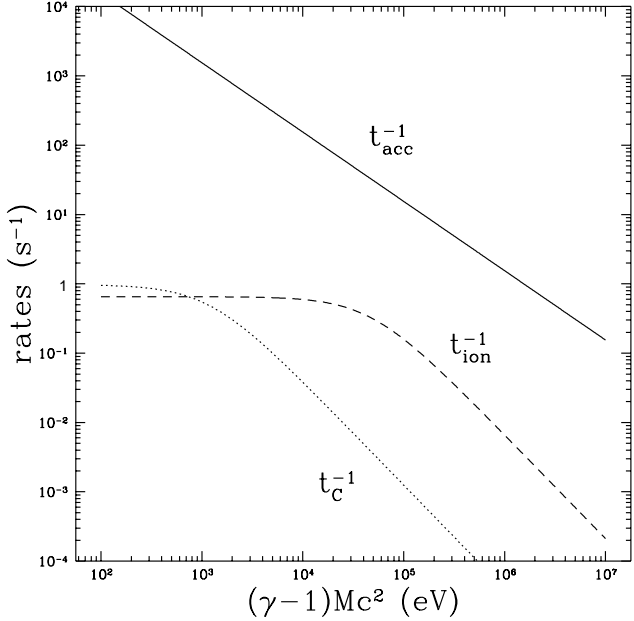}
 \caption{
The loss rates for ionisation ($t^{-1}_{\rm ion}$: dashed line) and Coulomb
($t^{-1}_{\rm C}$: dotted line) losses compared to the acceleration rate
for diffusive shock acceleration ($t^{-1}_{\rm acc}$: solid line). 
Curves are plotted for parameters appropriate to 
a supernova occurring in a dense stellar wind: 
$n_{\rm H}=10^9\,{\rm cm^{-3}}$, $x_{\rm i}=0.99$,
$B=1\,{\rm mG}$. The spatial diffusion coefficient
upstream and downstream of the shock is assumed equal
to the Bohm value. The shock speed is $10^4\,{\rm km\,s^{-1}}$ 
and the compression ratio is 
taken to be 4. }
\label{lossesfig}
\end{figure}

The Coulomb and ionisation losses are roughly constant for particles of speed below
$\beta_{\rm th}c$ and $\beta_0 c$ respectively and 
then fall as $\beta^{-3}$ at higher velocities, 
see Fig.~\ref{lossesfig}. Comparing this behaviour with that of $\accrate$ it
is clear that the ionisation and Coulomb losses will not suppress the shock 
acceleration at any energy if the acceleration exceeds the loss rate 
for particles of speed 
$\beta_{\rm th}c$ and $\beta_0 c$ respectively. This is the case if
\eqb
 && k_1^{-1} 
\left(u_1\over 10^3\,{\rm km\,s^{-1}}\right)^2
\left(B\over 1\,\mu{\rm G}\right)
{1\over n_{\rm H}}
\nonumber\\
&\gg& 10^{-6}{\rm Max}\left[x_{\rm i}\Tfour^{-1/2},(1-x_\mathrm{i})\right]
\enspace.
\eqe
Provided particles are injected at a speed several times the 
ion thermal velocity, so that the distribution function can
be approximately isotropic in both the upstream and downstream 
frames (Kirk \& Schneider~\cite{kirkschneider89}; 
Malkov \& V\"olk~\cite{malkovvoelk95}) and the self-excited waves are not
damped by thermal ions, we can expect that $k_1\gesim1$. In this
case, the constraint is not restrictive for the interstellar medium.

\section{Conclusions and implications for observations}

With respect to observational tests of shock acceleration in supernova
remnants, the limit imposed by upstream wave damping on the maximum
particle energies is significant, but not serious, for the
observability of supernova remnants in gamma-rays using the
atmospheric Cherenkov technique.  The limit can of course be
circumvented by locating the particle acceleration and the gamma-ray
production target in different regions or phases. In a clumpy medium,
one could imagine accelerating in the low-density interclump phase
with the dense clumps behind the shock providing the target
material. Or one can consider the possibility of a SNR exploding near
a molecular cloud (Aharonian et al.~\cite{aharoniandruryvoelk94}).  
In addition one should consider
the pre-ionisation of the upstream medium by the soft X-ray and UV
radiation from behind the shock, an effect which will also reduce the
effect of ion-neutral damping.

One other possibility for obtaining observational evidence for shock
acceleration in SNRs is the use of Balmer diagnostics in non-radiative
shocks (Raymond~\cite{raymond91}; Smith et al.~\cite{SRL94}) 
to probe the structure of the cosmic ray
precursor. This technique essentially determines the upstream plasma
temperature and velocity averaged over the ion-neutral charge exchange
length 
\eqb
L_{\rm exch} = U {1\over n_{\rm i}\left<\sigma v\right>}
\eqe
which is numerically
\eqb
4\times 10^{-3}
\left(U\over 10^3\,\rm km\,s^{-1}\right)
\left(n_{\rm i}\over 1\,{\rm cm^{-3}}\right)^{-1}
\left(T\over 10^4\,\rm K\right)^{-0.4}\,{\rm pc}.
\eqe
This has to be compared with the characteristic scale associated with
particles of energy $E$, $\kappa_{\rm B}(E)/U$. The two are equal at
\eqb
\left(E\over 1\,\rm TeV\right) &=&
0.04 
\left(U\over 10^3\,\rm km\,s^{-1}\right)^2
\left(n_{\rm i}\over 1\,{\rm cm^{-3}}\right)^{-1}\nonumber\\
&\times&
\left(T\over 10^4\,\rm K\right)^{-0.4}
\left(B\over 1\mathrm{\mu G}\right).
\eqe
The balmer diagnostic technique can in principle detect that part of
the precursor structure produced by particles above this energy. 

In conclusion, the main result of this paper is that a more detailed
analysis of the constraints on particle acceleration in dense media
does not support the pessimistic view expressed by Draine and McKee~(\cite{DM}) 
``shocks that are optically observable, either as
nonradiative shocks in a partially neutral medium or as radiative
shocks, are generally unable to accelerate particles to extremely
relativistic energies''. As far as the constraints imposed by upstream
ion-neutral wave damping, and ionisation and Coulomb losses are
concerned optically visible SNR shocks should be capable of
accelerating cosmic rays either to the revised cutoff
(\ref{numcutoff}), or, if the condition (\ref{nolimit}) is satisfied
to the maximum energy allowed by the shock geometry and age.

\section*{Acknowledgement}

This work was supported by the Commission of the European Communities
under HCM contract ERBCHRXCT940604.


\begin{thebibliography}{}

\bibitem[1986]{achterbergblandford86}
Achterberg A., Blandford R.D., 1986, MNRAS, 218, 551

\bibitem[1994]{aharoniandruryvoelk94}
Aharonian F.A., Drury L.O'C., V\"olk H.J. 1994 A\&A 285, 645

\bibitem[1978]{bell78}
Bell A. R. 1978, MNRAS, 182, 147

\bibitem[1988]{BK88}
Berezhko E. G., Krymsky G. F., 1988, Sov.\ Phys.\ Usp., 
31, 27

\bibitem[1987]{BE87}
Blandford R. D., Eichler D., 1987,
Physics Reports, 154, 1

\bibitem[1993]{DM}
Draine B. T., McKee C. F., 1993, ARAA, 31, 373

\bibitem[1991]{JE91}
Jones F. C., Ellison D. C., 1991, Space Sci.\ Rev.\ 58, 259

\bibitem[1989]{kirkschneider89}
Kirk J. G. \& Schneider P. 1989 A\&A 225, 559

\bibitem[1971]{KC71}
Kulsrud R., Cesarsky C. J., 1971, ApL, 8, 189

\bibitem[1969]{KP69}
Kulsrud R. M., Pearce W,. 1969, ApJ, 156, 445

\bibitem[1983]{lagagecesarsky83}
Lagage P. O., Cesarsky C. J., 1983 A\&A 118, 223  

\bibitem[1995]{malkovvoelk95}
Malkov M.A., V\"olk H.J. 1995 A\&A 300, 605

\bibitem[1994]{mannheimschlickeiser94}
Mannheim K., Schlickeiser, R. 1994 A\&A 286, 983

\bibitem[1991]{raymond91}
Raymond J. C., 1991, PASP, 103, 781

\bibitem[1994]{SRL94}
Smith R. C., Raymond J. C., Laming J. M., 1994, ApJ, 420, 286

\bibitem[1981]{VMF81}
V\"olk H. J., Morfill G. E., Forman M. A., 1981, ApJ, 249, 161

\end{thebibliography}
\end{document}